# Catalysis of Transmutations by Heavy Electron Quasiparticles in Crystallites

Anthony Zuppero[1], Thomas J Dolan[2],

1 Tionesta Applied Research Corporation, Sequim, WA
2 University of Illinois at Urbana-Champaign

## Contents

## Abstract

This article describes our hypothesis on how transmutations may be induced by solid state effects in a crystalline lattice. We discuss the chemical reaction case, our extension to the nuclear binding case, and a tri-body model of a heavy electron quasiparticle catalyzing the binding of two nearby ions. For a given primary reaction we can estimate the required electron mass threshold m*, identify possible reaction products, estimate tunneling probabilities, and calculate energies available for each path. We compare model predictions with experimental data of transmutations, and consider hazards associated with experiments.

## Background

Scientists have been reporting nuclear transmutations and anomalous energy generation accompanying chemical phenomena for about 30 years. Hydrogen and deuterium apparently can enable nuclear transmutations of reactants that generate elements and isotopes that were not originally present. The isotopes provided the key clues to understand the process. The energy generated exceeded that available from chemical reactions, inspiring worldwide interest. Storms (2014) provides abundant background information, and Biberian (2020) describes the current state of research.

Our hypothesis derives from a chemical physics binding reaction discovered during the 2000s. Researchers at UC Berkeley and UC Santa Barbara discovered a binding reaction referred to as “Vibrationally Promoted Electron Emission“ (VPEE) by LaRue et al., “chemicurrent” by Nienhauss et al., and “nanodiode” by Somorjai, Ji, Zuppero, Gidwani, et al.

LaRue (2011) documented that when reactants attracted to electrons between them begin as almost totally separated entities, and when the electron effective mass satisfies a simple function of binding and coulomb energy and reaction product size, the tri-body is unstable and promptly collapses and binds. The binding energy becomes partitioned into the kinetic energy of a liberated, ejected electron and the internal vibration of the reaction product.

This discovery quantified the condition where a certain type of tri-body reaction undergoes a very fast and prompt binding, and the energy can be all in the electron, all in the reactant vibration, or mixed.

For example, in the chemical physics case a highly vibrationally excited molecule may attract an electron from a metal surface, transfer most of the vibrational energy to the electron, and then eject it.  Or CO may bind to O on a conducting surface, ejecting an electron with the binding energy. (Zuppero and Dolan 2009)  We postulate that a similar reaction may also occur on the nuclear scale, which can have a similar potential energy curve.  We substitute nuclear binding potentials for chemical ones and add known solid-state, heavy electron physics.

## Transmutation Observations

Some reactions appear to bind pairs of protons or deuterons (or other light nuclei) to heavier reactants as if catalyzed on the surface of a crystalline lattice. The reactants include Ca, Ti, Ni, Sr, Pd, Cs, Ba, W, and U and radioactive isotopes. The low-mass isotopes may include hydrogen, deuterium, tritium, Li, Na, and K. Only trace neutrons and gamma rays are observed. Almost the entire periodic table could apparently take part in electron quasiparticle-catalyzed nuclear transmutations in appropriate lattice conditions.

One of the reactant ions may have any number of positive charges. Molecular reactants can bind chemically due to a *chemical* binding potential. Nuclear reactants can bind together with a *nuclear* binding potential, if the tri-body can contract to a size  close enough to the nuclear force radius for binding to occur. To contract to this size, electron tunneling must occur.

## Quantum Kinetic Energy of Confinement

We use a one-dimensional tri-body model, with an electron quasiparticle and a molecular or nuclear binding potential between a single-positive and multiple positively-charged reactants. Even though the net coulomb attraction in solid materials is attractive, and there are no net repulsive forces, the electron’s quantum kinetic energy of confinement (QKEC) from the Heisenberg Uncertainty Principle (HUP) provides a repulsive momentum to resist collapse of everything to nuclear density.

The electron energy associated with confining an electron to a region is the quantum kinetic energy of confinement QKEC.[Ashkenazi 2006].  Kinetic energy is represented by

$$\mathrm{QKEC} = \langle (p-p_0)^2/2m^* \rangle = \sigma^2_p/2m^*$$

where p is the momentum and m* is the effective mass.

The ion separation is characterized by its variance $\sigma^2_x$, which is a measure of the size of the electron confinement region $\sigma^2_x = \langle(x-x_0)^2\rangle$.

The Robertson- Schrődinger relation (modern Heisenberg relation) relates variances of momentum and position (relative to center of mass $x_0$ and momentum $p_0$).

$$\sigma^2_x\,\sigma^2_p = (\hbar/2)^2\,K(n) \quad \rightarrow \quad QKEC = (\hbar/2)^2\,K(n)/2m{*}\sigma^2_x$$

where $\hbar$ is the reduced Planck constant and K(n) is a form factor obtained by solving the Schrődinger equation.

Figure 1 shows the potential vs. ion separation distance $\sigma_x$ for a tri-body including attracting reactants and a normal electron. The curve below the axis is negative (coulomb attraction) and the dotted curve above the axis is repulsive (HUP).

How can we use a solid-state electron quasiparticle and thereby take advantage of an elevated effective mass? The condition for validity of the electron's effective mass demands that the electron be "non-interacting."

In a conducting solid, electrons and reactants have quantum expected positions at the equilibrium regions (equal forces in all directions), even though the equilibrium is unstable. This position is non-interacting.

A sufficiently heavy electron sluggishly moves in response to nearly equal forces in opposing directions. If the two reactants approach to within the force of their binding potential before the electron (or muon) moves away, they may promptly bind and share the energy with the third body. If, instead, the third body moves away first, the coulomb collapse fails and no reaction occurs.

Quantum mechanics readily approximates these two situations, as explained by comparing Figure 1 and Figure 2. Figure 1 shows a state with normal electron mass, where electron tunneling always finds no stable states at small, $\sigma_x$ binding dimensions. The large increase of QKEC (dotted line) at small $\sigma_x$ overwhelms the binding potential, preventing a reaction. A heavier electron mass can reduce the QKEC and yield the diagram of Figure 2, where tunneling can occur.

**Quantum kinetic energy of confinement**

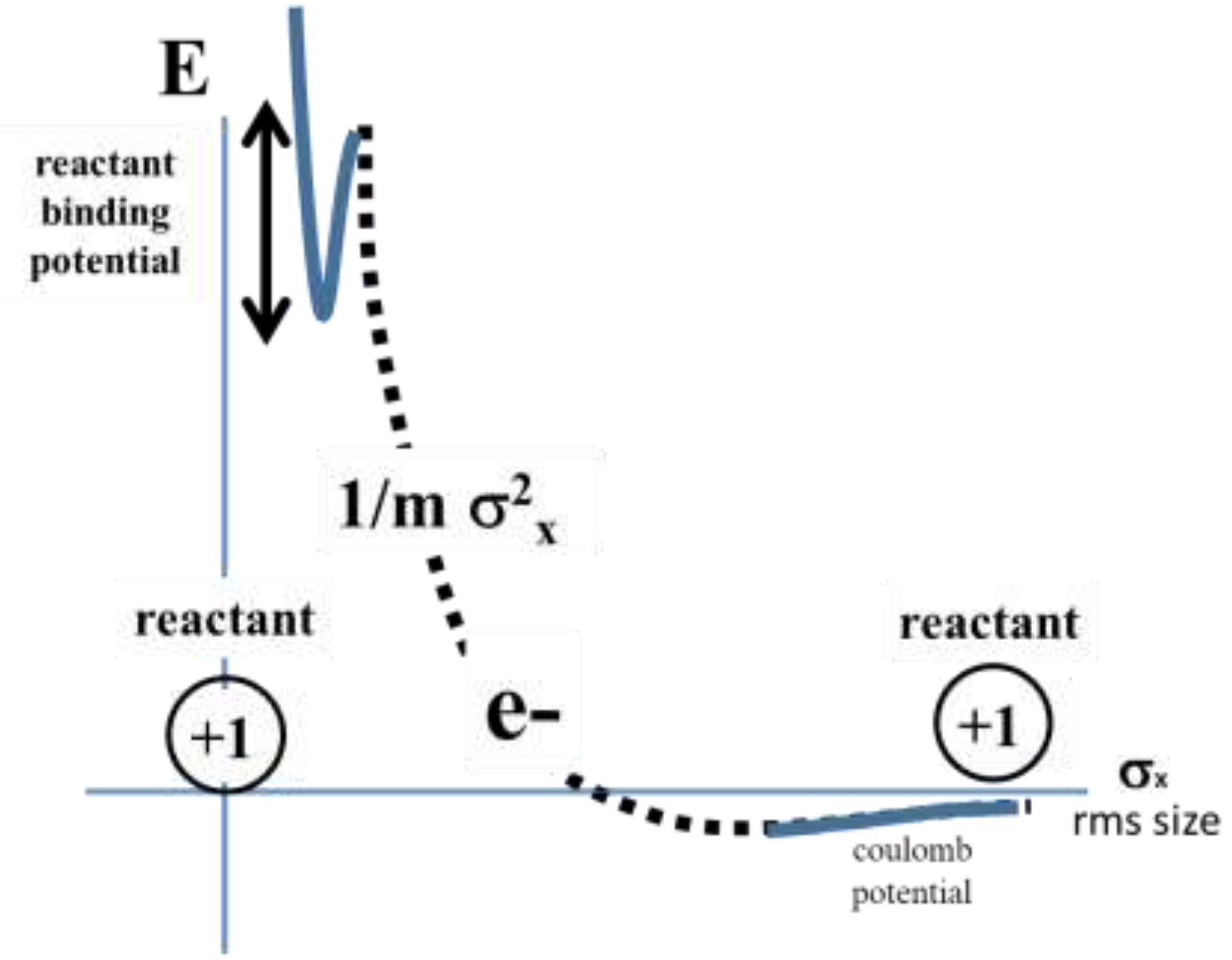

**Figure 1.** Potential curve with normal electrons. The total QKEC is about 200 MeV repulsive in the nuclear region, much larger than the reaction's nuclear binding potential ~ 3-12 MeV. Therefore, binding with normal electrons has vanishing probability.

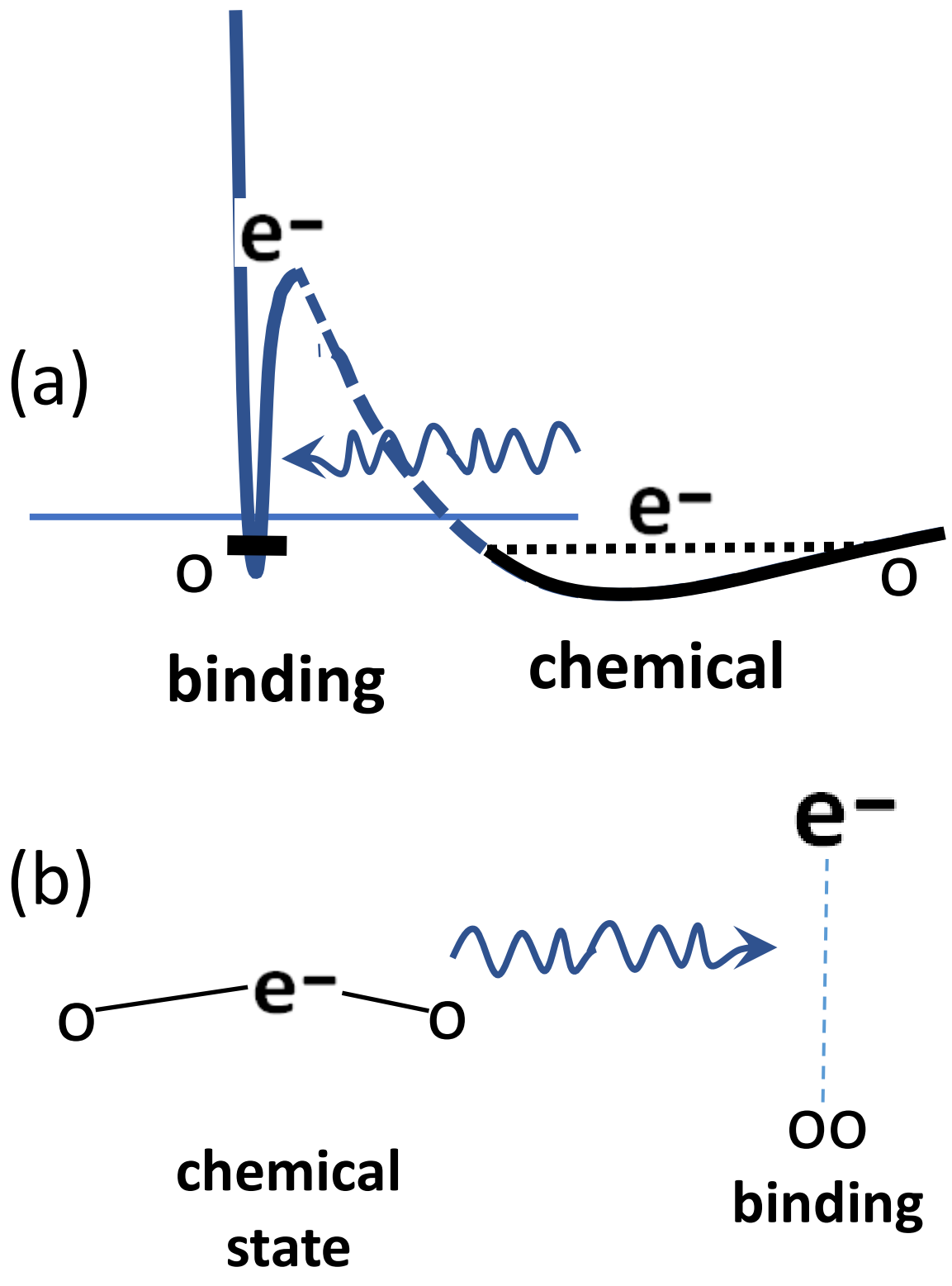

**Figure 2.** Potential vs. ion separation distance $\sigma_x$ for a case with a heavy electron quasiparticle. The tri-body comprises electrons (e-) and the reactant ions (o) attracted together in a three-body state initially of chemical size, (a). If he electron mass is heavy enough, the electron can tunnel to a three body state of binding size. The reactants are already in a state of coulomb attraction. Tunneling places the electron within the range of the binding potential well at nuclear force radius,(b). Tunneling results in prompt coulomb collapse and reactant binding. For electron-stimulated transmutations, the binding potential is nuclear, and the tunneling is to the nuclear force radius.

If the electron quasiparticle has a heavy effective mass, then the QKEC is reduced, and the nuclear binding potential may be low enough to form a stable potential well in the nuclear region, Figure 2. Then electron tunneling through the QKEC barrier becomes feasible. At threshold, the potential well of Figure 2a dips below zero, and is therefore stable. Prompt tunneling may occur, resulting in bound nuclear states, with third bodies redistributing the excess binding energy.

## Effective Mass Threshold

If the attractive forces of coulomb potential plus binding energy can exceed the QKEC at the nuclear force radius, then a stable potential well forms, and a nuclear binding reaction is possible.

$$(\text{binding energy} + \text{coulomb energy}) \geq (\hbar/2)^2\ K(n)/2m^*\sigma^2_x$$

at $\sigma_x$ = nuclear force radius $r_n$ (several fm). Solving for the threshold mass m* gives

$$m^* \geq (\hbar/2)^2\ K(n)\ /\ 2(\text{binding energy} + \text{coulomb energy})r_n^2$$

Methods for producing the heavy electron quasiparticles are discussed next. The estimates of threshold m* range from about 9 $m_o$ for deuterium reactions up to about 90 $m_o$ for hydrogen reactions ($m_o$ is the normal electron rest mass in a vacuum).

## Generating Heavy Electron Quasiparticles

The effective mass of an electron quasiparticle in solid state physics is

$$m^* = \hbar^2/(\partial^2E/\partial k^2)$$

where E is energy, k is crystal momentum in the band structure diagram, and ħ is the reduced Planck constant.(Kittel 2005) Heavy electron quasiparticles can be generated by injecting crystal momentum k and energy E into a crystallite lattice to place some electrons near inflection points of the band diagram, Figure 3.

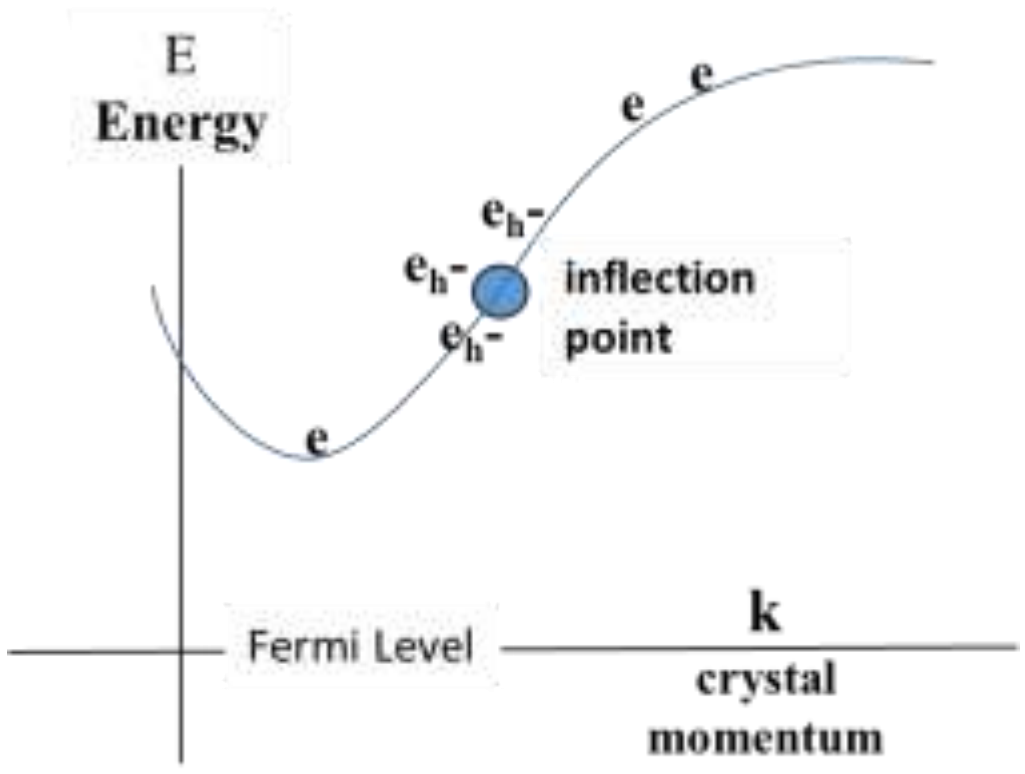

**Figure 3**. Energy vs crystal momentum in a hypothetical band structure diagram of a solid crystal, shown in the first Brillouin Zone, BZ 1.

In some lattices thermal energy is sufficient. Crystal momentum may be injected by many means, as illustrated in Figure 4.

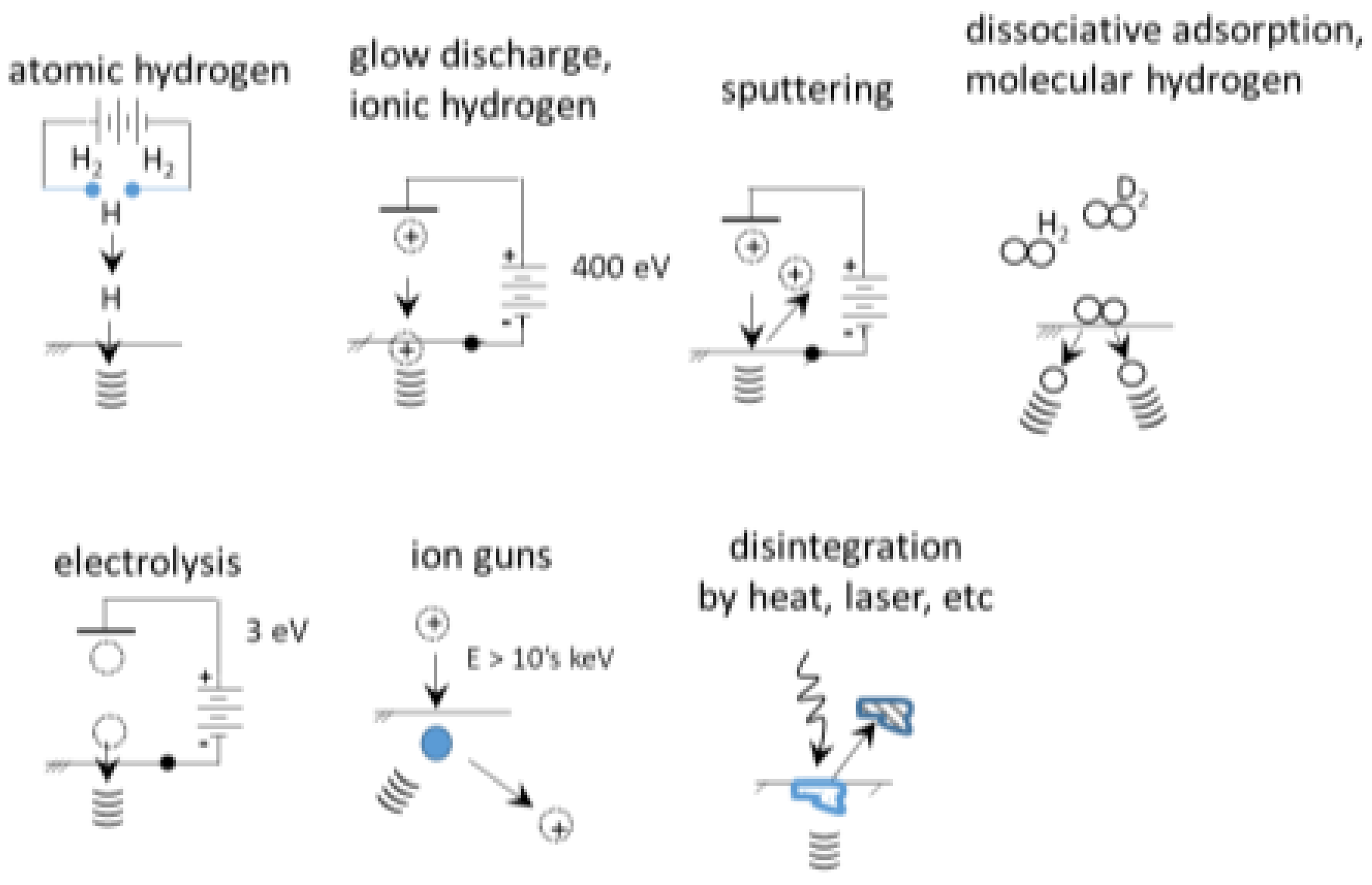

**Figure 4.** Methods of crystal momentum injection.

All the observed transmutations known to us have a momentum-injection trigger.
Even clumsy injection results in a spread of crystal momenta and electron energies near inflection points, where some elevated effective mass electron quasiparticles are created. Figure 5 shows an inflection point in the band structure curve. Electron quasiparticle effective mass is proportional to the inverse of the curvature, which becomes very large at the zero-curvature inflection point.

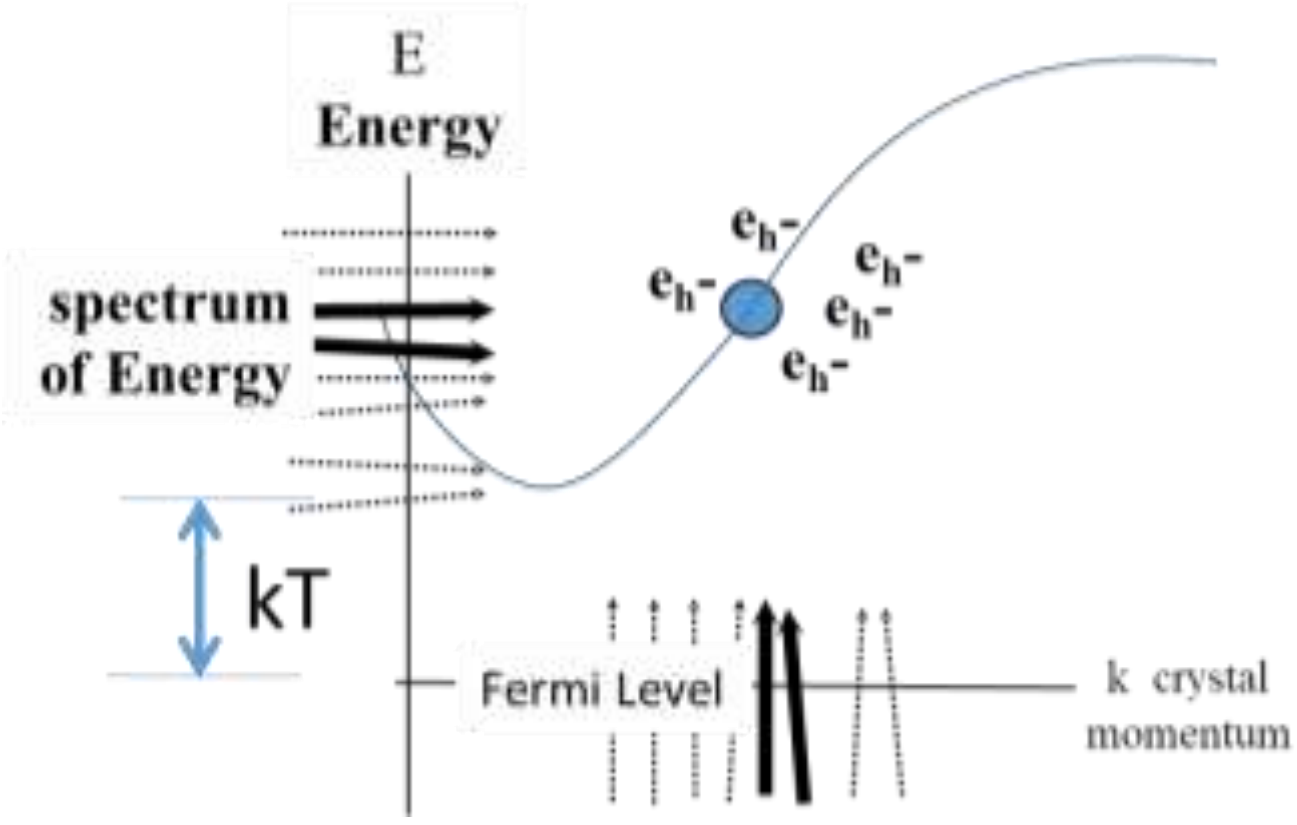

**Figure 5**. Spreads of energy and momentum injection into the first Brillouin Zone.

The crystal momentum wavelength must be short enough to access inflection points in the first Brillouin Zone of the crystallite region. “Short” means a wavelength of no more than several times the unit crystal dimension. All the stimulation methods of Figure 4 may be tailored to satisfy this condition.

## Tunneling

The reacting system must tunnel to state whose size $\sigma_x$ is small enough for nuclear binding forces to have effect. The collapse must occur before the electron interacts significantly (collides). The simplest case, evaluated here, starts with an electron *exactly* at the equilibrium point. The collapse occurs to inside the nucleus.

When the electron collides with the reactants approaching it from opposite sides it can no longer use the lattice resonating property to provide an effective mass. It responds like a normal electron and scatters.

In the *chemical* counterpart reaction, the electron is ejected and shares the binding energy with the product nucleus, left in a vibration state. However, in the *nuclear* transmutation case, an ejected electron with the binding energy (~ 6 – 12 MeV) has not been observed. Only the expected product in its ground state is observed.

With a heavy electron quasiparticle the chemical separation distance is shrunk by the square root of the mass ratio $(m^*/m_o)^{1/2}$ We call the tunneling region inside the minimum chemical separation the "halo" region, which may be on the order of 3-10 nuclear force radii. One can associate this state with a halo nucleus excited to the point of dissociation, a Rydberg nucleus. This state is created by the reactants converging on N pairs of heavy electrons. The binding energy would be sufficient to dissemble the product into the original reactants. The compound nucleus in the halo region often fissions into stable fragments, such as He.

The portion of the electron wave function reaching the nuclear radius provides transient shielding that enables merger of reactant nuclei. Prompt coulomb collapse can occur if one or more additional bodies can absorb binding energy, also conserving momentum and spin; or, if the reactants fracture into fission products.

The tunneling probability is estimated as

$$P = \exp(-2G)$$

where the Gamow factor is

$$G = (2m)^{1/2} \int_a^b [E(x) - Eo]^{1/2} dx/\hbar$$

x represents $\sigma_x$, a is the nuclear force radius, and b is the chemical separation distance. This is a three-body tunneling at thermal conditions, not two-body tunneling, as in hot fusion. [Zuppero and Dolan 2019]

## $H_2$ and $D_2$ Pairs

A review of 30 years of isotope data shows that if we use pairs, for example, N pairs of protons, deuterons, or tritons, we can account for almost all the isotopes observed.

N pairs + central reactant + heavy electrons →
transmuted nuclei + internal energy + dissociation products.

We use the notation $H_2$ to represent a pair of H atoms in the lattice, keeping in mind that this is *not* a gas molecule. An example reaction is

$H_2$ + nickel + 2 heavy electrons ($m^* > 35m_o$) → Zn

is illustrated in Figure 6. The Zn may then fission into various branches (discussed later).

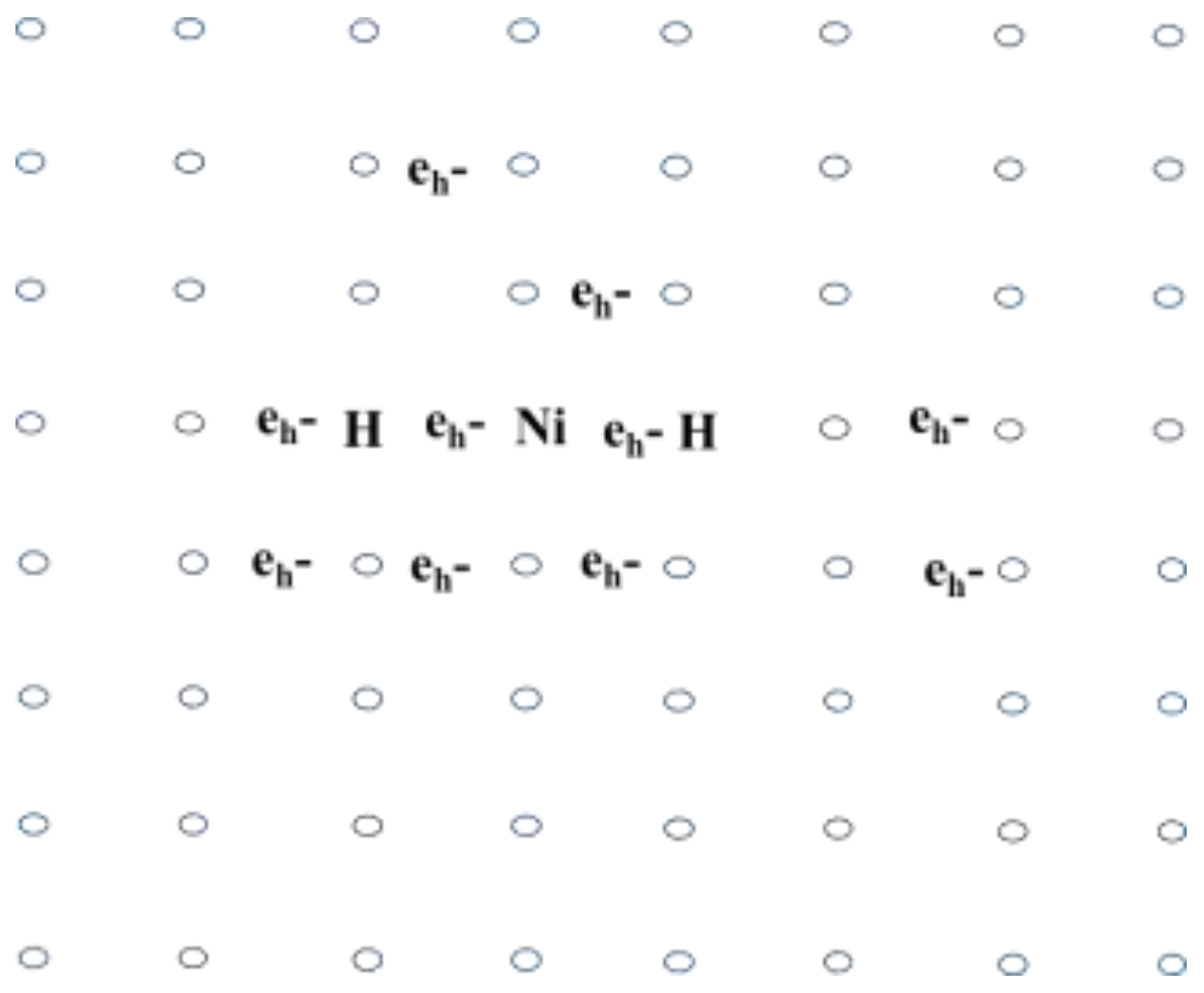

**Figure 6.** Ni reactants, H atoms, and heavy electron quasiparticles ($e_h$-) in a crystalline lattice serving as reaction region (blue circles). Normal electrons are not shown.

Electron quasiparticles experience *on average* equal forces in all directions. The quantum expected position values in the crystallite approximate "non-interacting" particles. A conducting crystallite approximates a resonant chamber (phonon lifetime ~ 3 ps) for the expected ballistic lifetime of the electron quasiparticle (~ 10 fs), so the phonon lifetime is long enough to support many generations of heavy electron quasiparticles. The duration of the heavy mass approximation must be at least long enough for reactants to undergo coulomb collapse to binding. Estimates show both the chemical and the nuclear transmutation cases have sufficient duration for neighbors to tunnel and bind
.

Are the heavy electron quasiparticles paired (spin up – spin down) ?

Figure 7 lists some transmutations that have been reported.

**N $D_2$ pair reactions Observed**

Iwamura et al. with Pd (gas adsorption)

$1D_2 + Ca \rightarrow Ti$

$1D_2 + W \rightarrow Os.$

$2D_2 + Cs \rightarrow Pr$ **50.6 MeV**

$2D_2 + Sr + \rightarrow Mo$

$2D_2 + W \rightarrow Pt$

$3D_2 + Ba \rightarrow Sm$

Dash et al. with Ti (electrolysis)

$1D_2 + Ti \rightarrow Cr$

$2D_2 + Ti \rightarrow Fe$

$3D_2 + Ti + \rightarrow Ni$

$4D_2 + Ti \rightarrow Zn$

**N $H_2$ pair reactions Observed**

$1H_2 + Ni \rightarrow Zn$ gas adsorption

$1H_2 + Rb \rightarrow {}^{87}Y$ (radioactive) electrolysis

**Figure 7.** The N pairs of hydrogen isotopes bind with a central reactant to form observed products. The estimated threshold effective mass for $D_2$ reactions is $m^* \sim 10\ m_o$ , and for $H_2$ reactions $m^* \sim 35\ m_o$. For simplicity the heavy electrons are not shown explicitly in the equations.

Various paths are possible for the compound nucleus that conserve hadrons, energy, momentum, and spin. Reaction paths may include emission of energetic electrons, neutrinos, gamma rays, or x-rays; internal excitations; and fission (fracturing into stable fragments, such as He). For a given primary reaction we can estimate the required mass threshold m*, identify possible reaction products, estimate tunneling probabilities, and calculate energies available for each path. We compare model predictions with experimental data of transmutations. We are not able to compute reaction rates and branching ratios.

## Fission Products

The collapsing reactants and electron quasiparticles have an approximately Rydberg state with chemical size greatly reduced by heavy electron screening. Collisions may terminate an electron's heavy inertia, but the three-body size may already be small enough to facilitate tunneling the rest of the way to nuclear dimensions, forming a "halo" (cloud of particles like a compound nucleus). The nucleons arrange into stable fragments that may promptly fission. We can estimate which fracture products will be stable.

N pairs of low mass isotopes are catalyzed by sufficiently heavy electrons to cause a rearrangement of the collective set of nucleons in the N-body cloud. The fission products conserve total hadron numbers. For example

$H_2$ + nickel-62 $\rightarrow$ iron-56 + 2He + 3.6 MeV

has 30 protons and 34  neutrons.

There is typically enough energy to rearrange the nucleons in the halo region, resulting in the observed fission products. These reactions release about 6 MeV per proton or neutron taking part in binding. The binding energy of alpha particles is typically less than this for nuclei heavier than about iron, and it decreases to negative values for radioactive uranium and high-mass elements, as shown in Figure 8.

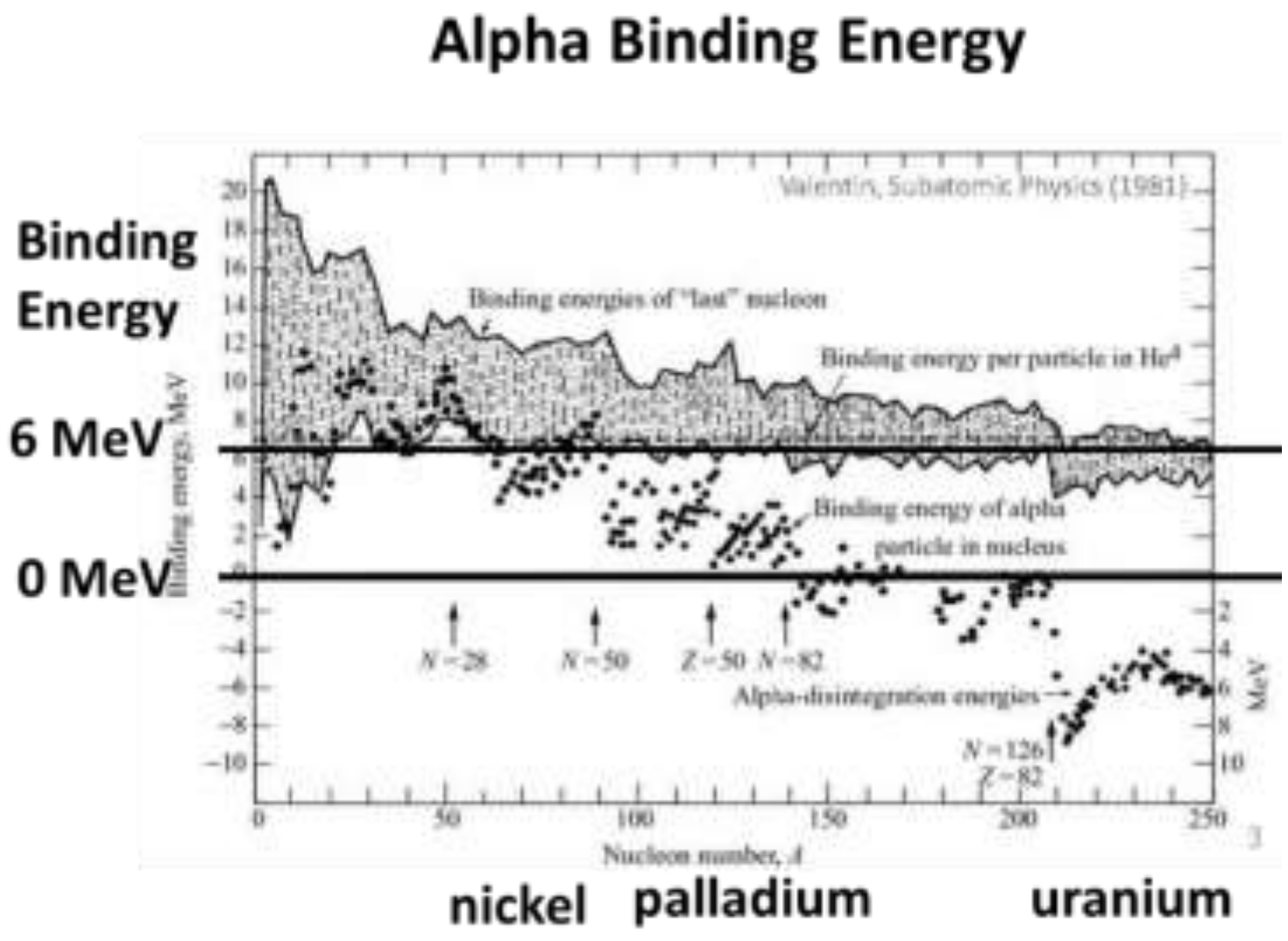

**Figure 8.** Alpha particle binding energies. (Valentin 1981)

## Palladium and $D_2$ Reaction Products

We postulate that some of the (formerly-heavy) normal electrons in the halo region attach to reaction products. This can produce energetic neutral helium and single-charged helium, as well as neutral hydrogen, deuterium, and carbon atoms. Data suggest this happens routinely, and that this energetic neutral emission could be the dominant energy path. Deuterium reactions could be dangerous, because the neutral helium product could be highly energetic (~23 MeV), highly penetrating, and difficult to detect.(Appendix A)

The reactants apparently form halo nuclei in a system with sufficient energy to cause fission. Figure 9 shows fission products formed using $ND_2$ fuels, measured by EDX spectroscopy.

| |
|---|
| **1 D2 +Pd--> Aluminum and Bromine** |
| **$1d_2$** +2 m*10 + $^{102}Pd$ --> $^{106}Cd$ 25.5 MeV --> **$^{27}_{13}Al$ + $^{79}_{35}Br$** 31.6 MeV |
| **$1d_2$** +2 m*10 + $^{104}Pd$ --> $^{108}Cd$ 27.3 MeV --> **$^{27}_{13}Al$ + $^{81}_{35}Br$** 32.0 MeV |
| |
| **2 D2 +Pd--> iron and chrome** |
| **2d2** +4 m*10 + $^{102}Pd$ --> $^{110}_{50}Sn$ 51.8 MeV --> **$^{56}_{26}Fe$ + $^{54}_{24}Cr$** 82.1 MeV |
| **2d2** +4 m*10 + $^{102}Pd$ --> $^{110}_{50}Sn$ 51.8 MeV --> **$^{57}_{26}Fe$ + $^{53}_{24}Cr$** 80.1 MeV |
| **2d2** +4 m*10 + $^{104}Pd$ --> $^{112}_{50}Sn$ 51.8 MeV --> $^{58}_{26}Fe$ + $^{54}_{24}Cr$ 82.2 MeV |
| 2d2 +4 m*10 + $^{106}Pd$ --> $^{114}_{50}Sn$ 53.1 MeV --> $^{58}_{26}Fe$ + $^{56}_{24}Cr$ 80.7 MeV<br>$^{56}_{24}Cr$ 5.9 m (Radioactive)--> $^{56}Mn$ 2.5 hr (Radioactive)--> $^{56}Fe$ (stable 3 hrs later) |
| |
| **3 D2 +Pd--> chrome, nickel** |
| **$3d_2$** +6 m*11 + **$^{104}Pd$** --> $^{116}_{52}Te$ --> $^{64}_{28}Ni$ + $^{52}_{24}Cr$ 111.9 |
| **$3d_2$** +6 m*11 + **$^{104}Pd$** --> $^{116}_{52}Te$ --> $^{62}_{28}Ni$ + $^{54}_{24}Cr$ 113.0 |
| **$3d_2$** +6 m*10 + **$^{105}Pd$** --> $^{117}_{52}Te$ -> $^{64}_{28}Ni$ + $^{53}_{24}Cr$ 112.78 |
| **$3d_2$** +6 m*10 + **$^{106}Pd$** --> $^{118}_{52}Te$ --> **$^{64}_{28}Ni$ + $^{54}_{24}Cr$ 112.9 MeV** |
| |
| **4 D2 +Pd--> iron, nickel** |
| **$4d_2$** +8 m*10 + **$^{106}Pd$** --> $^{122}_{54}Xe$ --> **$^{58}_{26}Fe$ + $^{64}_{28}Ni$** 144.4 MeV |
| **P.A. Mosier-Boss / Journal of Condensed Matter Nuclear Science 13 (2014) 432–442** |

**Figure 9.** A sampling of fission products associated with $ND_2$ + Pd reactions, where an EDX spectrum indicates only the dominant isotopes. Experimental data are in boldface. (Mosier-Boss 2014)

(Heavy electrons are understood, but not listed explicitly in the equations hereafter.)

In summary: Pd + N deuterium pairs gives:

1 $D_2$ + Pd → Al + Br + 32 MeV
2 $D_2$ + Pd → Fe + Cr + 56 MeV
3 $D_2$ + Pd → Ni + Cr + 113 MeV
4 $D_2$ + Pd → Ni + Fe + 144 MeV

Our heavy electron catalysis model predicts all the products measured by EDX.

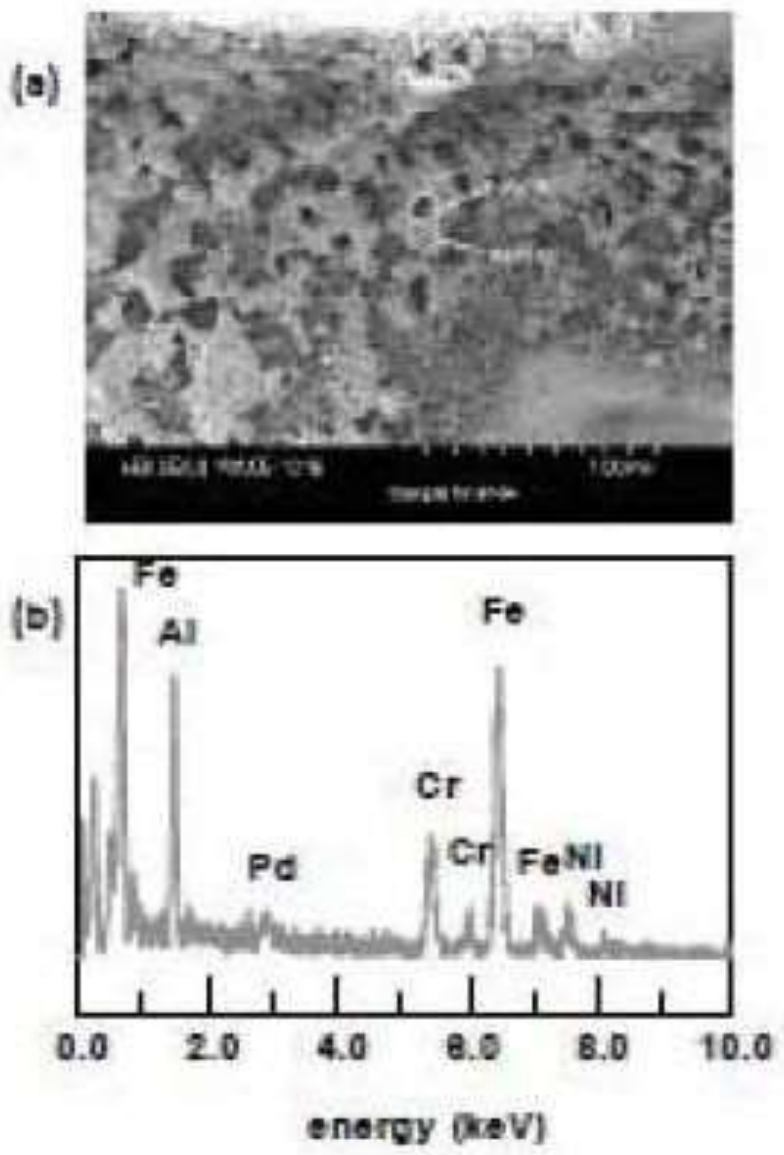

Figure 10. (a) SEM photomicrograph of the Pd deposit subjected to a magnetic field. (b) EDX analysis of one of the circled spots on the deposit. ( Mosier-Boss 2014)

This spectrum shows a line for aluminum, but not for bromine. This might be due to the fact that the bromine K-alpha is at 11.9 keV, and the EDX spectrum stops at 10.0 keV. The bromine L-alpha line lies directly under the aluminum line, so it is difficult to distinguish. Bromine is also a fuming liquid that may evaporate away quickly from the scan area.

Why chrome, iron, nickel, bromine and aluminum? They are the biggest sub-nuclei easily formed when one has a product formed with so much internal vibration energy that it will fission with no barrier. The available kinetic energy exceeds that needed for fracture products to escape. The excited compound nucleus is unstable to fission, as in the liquid drop model of uranium fission.

The halo fission reaction can emit helium and the original reactant, such as palladium. The halo alpha particle is expected to drag halo electrons with it, resulting in neutral helium emission. Our model postulates that

$$Pd + D_2 \rightarrow \text{(compound nucleus)} \rightarrow Pd + He .$$

The product looks like D+D → He two-body fusion, but it is not.

One fission product set includes the emission of helium, and suspected emission of carbon and oxygen, from various reactants R:

R + D2 → (X*) → helium + R
R + 2 D2 → (X*) → 2 helium + R
R + 3 D2 → (X*) → carbon + R
R + 4 D2 → (X*) → oxygen + R

The intermediate excited nucleus X* differs in each case. Helium, carbon, and oxygen have appeared as transmutation products. Many reactants R could fit in the above relationships. Heavy electron catalysis yields helium and the original reactant.

In some cases the ratio $^{108}$Pd/$^{110}$Pd becomes depleted, and other isotope ratios shift: "*The concentrations of* $^{109}$*Ag,* $^{59}$*Co, and* $^{64}$*Zn were also found significantly increased over the untreated palladium. This result is difficult to explain unless these elements were to result from fission after a deuteron was added to Pd*." (Storms 2014). Explanation is relatively simple using N pairs of deuterons .

## Tungsten and Nickel Reactions with $H_2$

Some reactions of $H_2$ with tungsten reactants are

W + $H_2$ → (Os) → fission products.

Figure 11 shows some expected fission product isotopes from the osmium compound nucleus.

$^{182}$W --> $^{36}$**S**_16 + $^{148}$**Nd**_60 74.4 MeV
$^{184}$W --> $^{48}$**Ca**_20 + $^{138}$**Ba**_56 101.3 MeV
$^{180}$W --> $^{58}$**Fe**_26 + $^{124}$**Sn**_50 115.3 MeV
$^{182}$W -- > $^{30}$**Si**_14 + $^{154}$**Sm**_62 63.3MeV
$^{182}$W --> $^{58}$**Fe**_26 + $^{124}$**Sn**_50 114.5 MeV
$^{182}$W --> $^{48}$**Ti**_22 + $^{136}$Xe_54 101.2 MeV
$^{180}$W --> $^{40}$Ar_18 + $^{140}$**Ce**_58 84.5 MeV
$^{182}$W --> $^{34}$S_16 + $^{150}$Nd_60 64.6 MeV

**Figure 11.** Some fission fragments expected from tungsten and hydrogen reacting to yield an osmium compound nucleus. Boldface letters indicate those claimed to be observed. Effective mass threshold $m^* = 22m_o$. (Mizuno 2005)

Figure 12 shows some expected fission products from the reaction

$$Ni + H_2 \rightarrow (Zn) \rightarrow \text{fission fragments} + \text{energy}$$

Neutral helium, hydrogen, and $He^+$ are expected, but may not be detected unless carefully sought.

| obscured | | observed | |
|---|---|---|---|
| $^4He+$ | $^4He+$ | $^{54}Fe$ | 1.5 MeV ($^{60}Ni$) |
| $^4He+$ | $^4He+$ | $^{56}Fe$ | 3.6 MeV ($^{62}Ni$) |
| $^4He+$ | $^4He+$ | $^{58}Fe$ | 4.8 MeV ($^{64}Ni$) |
| $^4He$ | p | $^{59}Co$ | 0.3 MeV |
| H | | $^{63}Cu$ | 6.1 MeV |
| $^4He$ | $^{A-2}Ni$ | | 9 MeV |

**Figure 12.** Obscured fission products are peculiar feature of transmutation fissions -- neutral alpha particles (helium), neutral hydrogen. Obscured single-charged helium are expected from reaction hydrogen and nickel with effective mass about 35 $m_0$. The observed fission fragments from hydrogen and nickel-60, -62, and -64 react to produce excited zinc compound nuclei. The number of hadrons in the combination Fe + $2He^+$ equals those of the input 2 hydrogen + nickel.

The bottom fission product reaction could be

$$^{62}Ni + H_2 + 2e \rightarrow (Zn) \rightarrow {}^{60}Ni + He\ (\text{neutral})$$

The ratio $^{62}Ni/^{60}Ni$ is depleted. Chemical physics ("DIMET" dissociation induced by multiple electronic transitions) suggests neutral fission fragment emissions. This helium is expected to be neutral because two electrons are in the halo upon its formation. The 9 MeV neutral helium is unlikely to be noticed, because it acts like 4 neutrons; and the nickel-60 is one of the dominant isotopes, masking new nickel-60 nuclei.

## Neutral Helium

McKubre reacted $D_2$ gas molecules in fine Pd powder and measured both the neutral He gas produced and the heat generated, as shown in Figure 13. That the average heat per reaction (~ 31 MeV) is 40% more than expected (~ 23 MeV) could be partially due to the other fissions, shown in Figure 9, with energies ranging from 30 to 100 MeV. Similar correlations have been measured for experiments using electrolysis.

Several researchers sought but did not find 32 MeV worth of alpha particles. Therefore, we suspect the helium may be neutral, as expected from DIMET neutral emission in chemical physics.

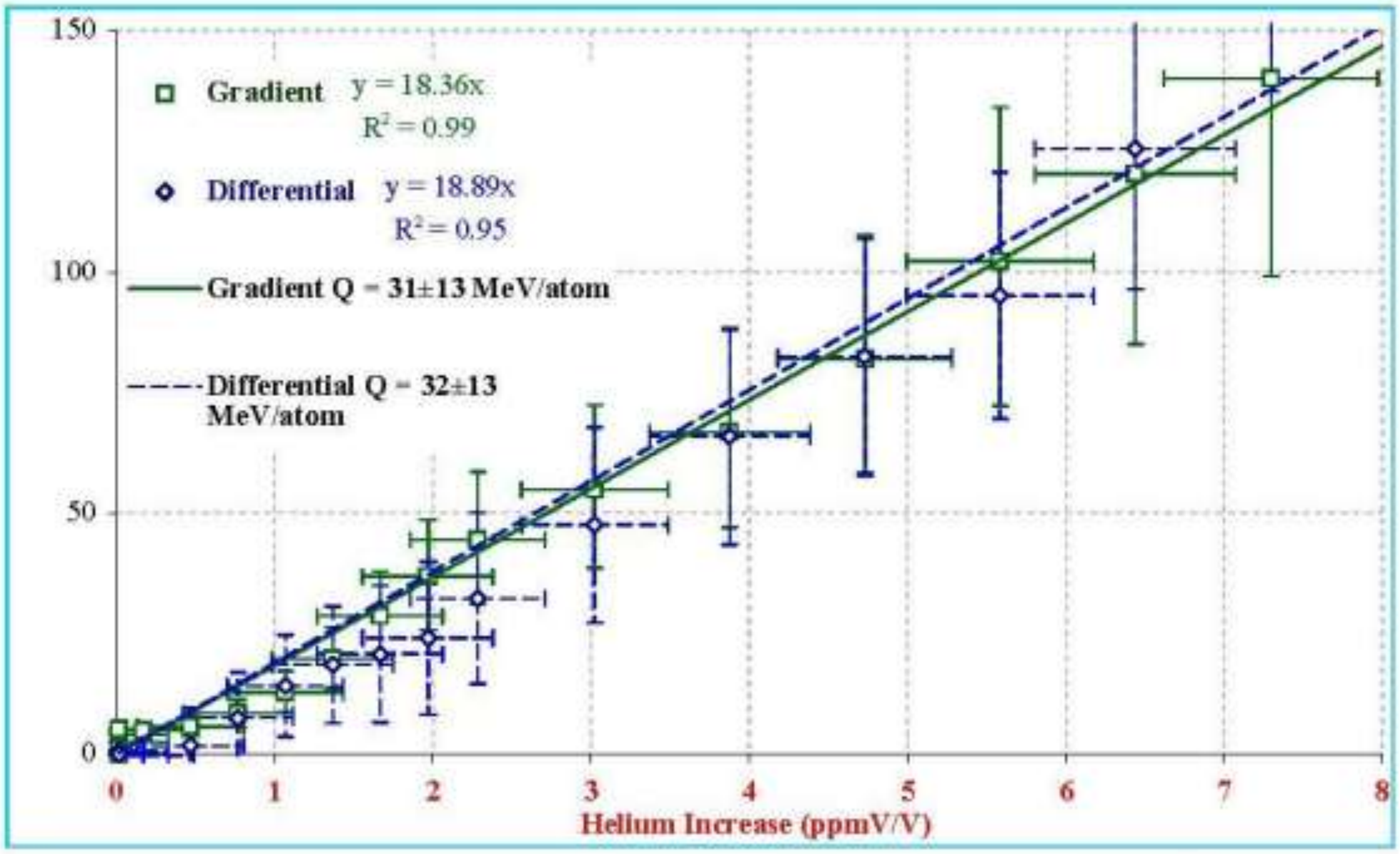

**Figure 13.** Correlation data between excess heat and $^4$He generation. (McKubre 2003 )

## Surface Reactions

Reactions have been demonstrated on the surface of selected crystallites made of elements that are different from the reactant. When Iwamura flowed deuterium gas through strontium and cesium chemical films deposited on a metal palladium surface, he observed the expected transmutations. (Iwamura 2003).

Bush (1994) used electrolysis with nickel in light-water and rubidium carbonate. He observed the expected production of strontium and an unexpected radioactive material with half life about 3.8 days, which is indicative of Yttrium-87 (3.35 d). These are consistent with fission of compound nuclei resulting from reactions of hydrogen with rubidium. The reaction was a rubidium salt, not an atom, on a nickel catalyst surface, not in it.

## Electron Catalysis Model

Our electron catalysis model suggests that most elements in the periodic table have an isotope that could react with ordinary hydrogen and emit MeV energies in appropriate lattice conditions. Attraction by heavy electron quasiparticles shrinks the ion-ion separation distance to where the electron can tunnel through the QKEC barrier to nuclear dimensions, screening the ion-ion collapse. When deuterium is the light reactant, most isotopes in the periodic table could react, if the electron mass were high enough.

Muon catalyzed fusion is a special case of our model. The negative mu meson (mass = $207m_o$) plays the role of the heavy electron quasiparticle, shrinking the dd molecule size, followed by tunneling through the QKEC barrier and re-emission of the muon. According to a physics approximation learned during the 1960s for muon catalyzed fusion, one can treat a system of heavy electrons amid a sea of normal electrons, a first approximation, as if the fuels and reactants were bare nuclei on the surface. Iwamura deposited a sub-monolayer of $Sr(OH)_2$ and the products fit the $ND_2$+reactant pattern. Bush deposited rubidium carbonate and the products fit the $NH_2$+reactant pattern.

This could be highly useful in continuous reactors. The fuels, reactants and ashes may flow into and out of a region of reaction crystallites whose surfaces host the reactions, when materials properties allow. For example, $Ca(OH)_2$ could be reacted with hydrogen to produce argon, trace titanium and scandium, and a pair of single-charged helium atoms (0.5 to 2 MeV).

## Possible Applications

Charged nuclei usually have too short a range to be easily detected, except for a pair of $He^+$ expected from one branch of hydrogen-pair-plus-reactant systems. The postulated (but not yet measured) copious emission of $He^+$ across a diode gap could result in direct electric power generation.

Energetic neutral atom emission could provide rocket thrust with high specific impulse.

Energetic charged particle emission could facilitate magnetic nozzle rocket propulsion. The postulated but not yet observed copious emission of single-charged helium nuclei could result specific velocities up to 2% the speed of light in a magnetic nozzle rocket propulsion, if free electrons were dragged along to neutralize the exhaust.

Stimulating pulsed reaction streams in propellant could provide energetic impulse against turbine blades. The range of the energy deposition by emitted ions is consistent with the size of the boundary layer (~ 1 mm) in turbine blades.

Appendix A describes micron-sized explosions that might indicate rapid reactions yielding bursts of high energy. Macroscopic explosions have destroyed equipment and injured people.

## Conclusions

Our tri-body catalysis model considers a slow electron quasiparticle between two reactant ions in a crystallite lattice. The attractive coulomb potential is opposed by the back pressure from confinement of the electron, due to the Heisenberg Uncertainty Principle, which prevents coulomb collapse. We assume that injection of energy and crystal momentum can bring the electron quasiparticle transiently near an inflection point of the band diagram, raising its effective mass and shrinking the ion separation. Then part of the electron evanescent wave function may tunnel to nuclear force dimensions (several fm), shielding merger of the two reactant ions. The electron is scattered when it collides with the converging reactant nuclei, and it is either ejected or it contributes to excitation energy of the compound nucleus, inducing fission.

Tri-body model predictions are consistent with experimental data:

- N pairs of deuterons ($ND_2$) reacting Ca, Cs, W, Sr, Ba, Ti, Ni, (Fig. 7)
- $ND_2$ reacting with Pd → (Al+Br), (Fe+Cr), (Ni+Cr), (Ni+Fe) (Fig. 9)
- Fission products of He, 2He, C, and O from $D_2$, $2D_2$, $3D_2$, and $4D_2$ respectively
- Fission products of Os resulting from reaction of W with $H_2$ (Fig. 11)
- Fission products of Zn resulting from reaction of Ni with $H_2$ (Fig. 12)

The helium and the isotopes have been observed. No two-body fusion need be invoked. Muon catalysis of hydrogen isotopes is another example of the tri-body reaction.

If we can understand and control these reactions, then some potential applications might become feasible: direct electric current generation, rocket thrusters, turbine thrusters, and applications of explosions.

Researchers should proceed with caution, because of the potential hazards associated with energetic neutral atoms and explosions (App. A).

---

# Appendix A. Hazards

## Energetic Neutral Atoms

The observed fission products for $ND_2$ + Pd are chrome, iron, nickel and predicted ***neutral*** helium, and must carry away 20 to 80 MeV. A 10 Watt source of neutral helium with 23.6 MeV is almost certainly dangerous. The range of a 23 MeV neutral He atom could be much longer than an alpha particle of the same energy. This results in dangerous, penetrating, nearly invisible radiation.

## Microscopic Explosions

Some reactants, such as Pd and $D_2O$, appear to trigger fast reaction rates. Volcano-like craters with 1-100 μm diameters suggest explosive energy release. Nagel and Srinivasan studied micro-explosions and emissions of sound bursts, radio-frequency waves, infrared, X-rays, neutrons, and charged particles. For example, sound bursts and tiny flashes of infrared light from the cathode surface occurred during co-deposition electrolysis. [Nagel 2014]

Figure 1 shows photos of various craters.

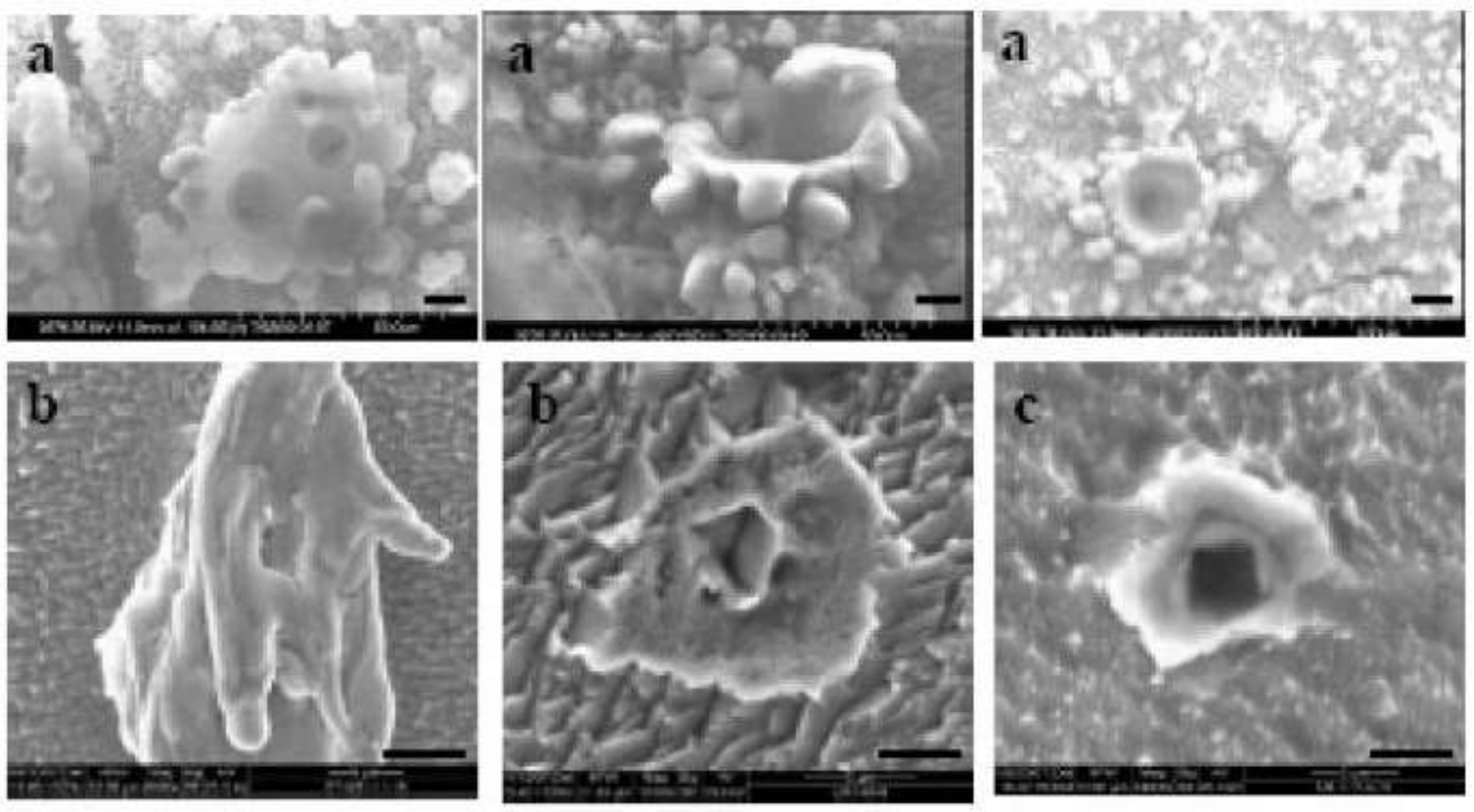

Figure 1. Craters from co-deposition experiments (top row) and from "super-wave" experiments (bottom row). [Nagel 2014]

The energy required to form these craters has been estimated by two methods:

- Calculating the volume of melted material and using the known volumetric energies of melting and vaporization.
- Scaling of energy releases from craters of many sizes, such as meteor impact, yielded an approximate equation for energy vs. crater diameter: Energy (J) ≈ 7.37 $D^{2.67}$, where D = crater diameter (m).

The estimates for energy required to melt or vaporize the material bracketed the size scaling equation. [Nagel 2013]

## Macroscopic Explosions

Some larger (centimeter size) meltdowns or explosions have damaged laboratories:

- **Fleischmann** and Pons reported meltdown of an electrolysis cell at the University of Utah in February 1985.

*They had one of their very first experiments set up in Room 1113 of the North Henry Eyring Building on the campus there at the University of Utah. They left it overnight and they came in in the morning and it was a mess. ... There was a [large] hole in the laboratory bench, there was a lot of particulate matter in the air.* [Rothwell 2007]

And they warned of danger:
*We have to report here that under the conditions of the last experiment, even using $D_2O$ alone, a substantial portion of the cathode fused (melting point 1554° C), part of it vapourised, and the cell and contents and a part of the fume cupboard housing the experiment were destroyed. ...Finally, we urge the use of extreme caution in such experiments: a plausible interpretation of the experiment using the Pd-cube electrode is in terms of ignition.* [Fleischmann 1989]

- T. P. **Radhakrishnan** reported an explosion at Bhabha Atomic Research Center, India, in September, 1989. An electrolysis cell was used to measure tritium production in $D_2O$. After many hours operation the electrolyte temperature "shot up" from 71°C to 80°C and the cell exploded. "*Later metallographic examination of the palladium cathode… showed an extensive twinning within the palladium grains with worm-like microstructure. This is suggestive of an intensive shock-wave impact on the metal".*
[Radhakrishnan 1989]

- X. **Zhang** reported explosions at the Southwest Institute of Physics, China, in April, 1991. Explosions occurred three times, blowing out the top plug or fracturing the cell, Figure 2.

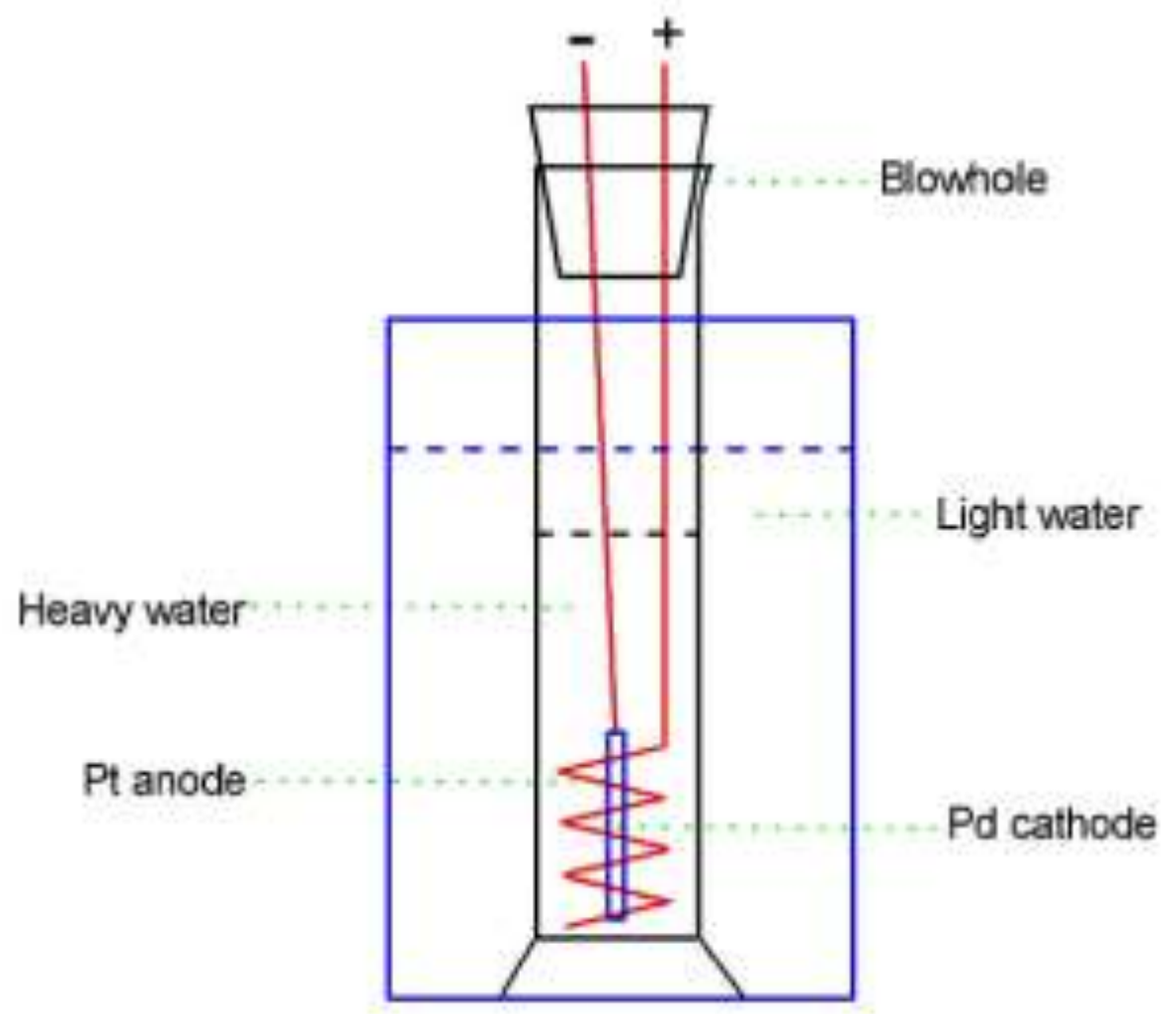

Figure 2. The electrolysis cell.

The system electrode current density was 62 mA $cm^{-2}$. After many hours of operation the power generated started rising. The authors state,

*After about ten seconds, the excess power was so great that the temperature at the nuclear reaction region (< 1 $mm^3$) of Pd tube rose to more than one thousand Celsius degrees, and the metal lattice distorted strongly as observed afterwards due to thermo-stresses; the surrounding heavy water vaporized, and the electrolyte boiled. … The excess power reached 5.1–5.5 kW. The explosion followed.*

About 12 kJ was released by the explosion, but only 0.31 kJ was available from hydrogen combustion. [Zhang 2015]

- An explosion occurred at **SRI International** January 2, 1992, killing Dr. Andrew Riley and wounding 3 others. On January 1 Dr. Riley corrected a leaky tube with a Swagelok fitting and returned the cell to the surrounding bath. On January 2, another leak was observed. *Dr. Riley first removed the clear acrylic top of the calorimeter and then lifted the calorimeter out of the water bath, set it on the edge of the bath, and was waiting for the water to drain back into the bath when the explosion occurred.* This explosion was attributed to combustion of deuterium and oxygen. It is not clear how a large quantity of these gases accumulated in spite of a recombiner. [Smedley 1992]

- Jean-Paul **Biberian** reported an explosion in Marseille, France, in September, 2004. An electrolysis cell (Figure 3) was run at various powers between 1 and 30 W for 350 hours, and then at about 0.7 W for about 400 hours, when an explosion destroyed the dewar, Figure 4.

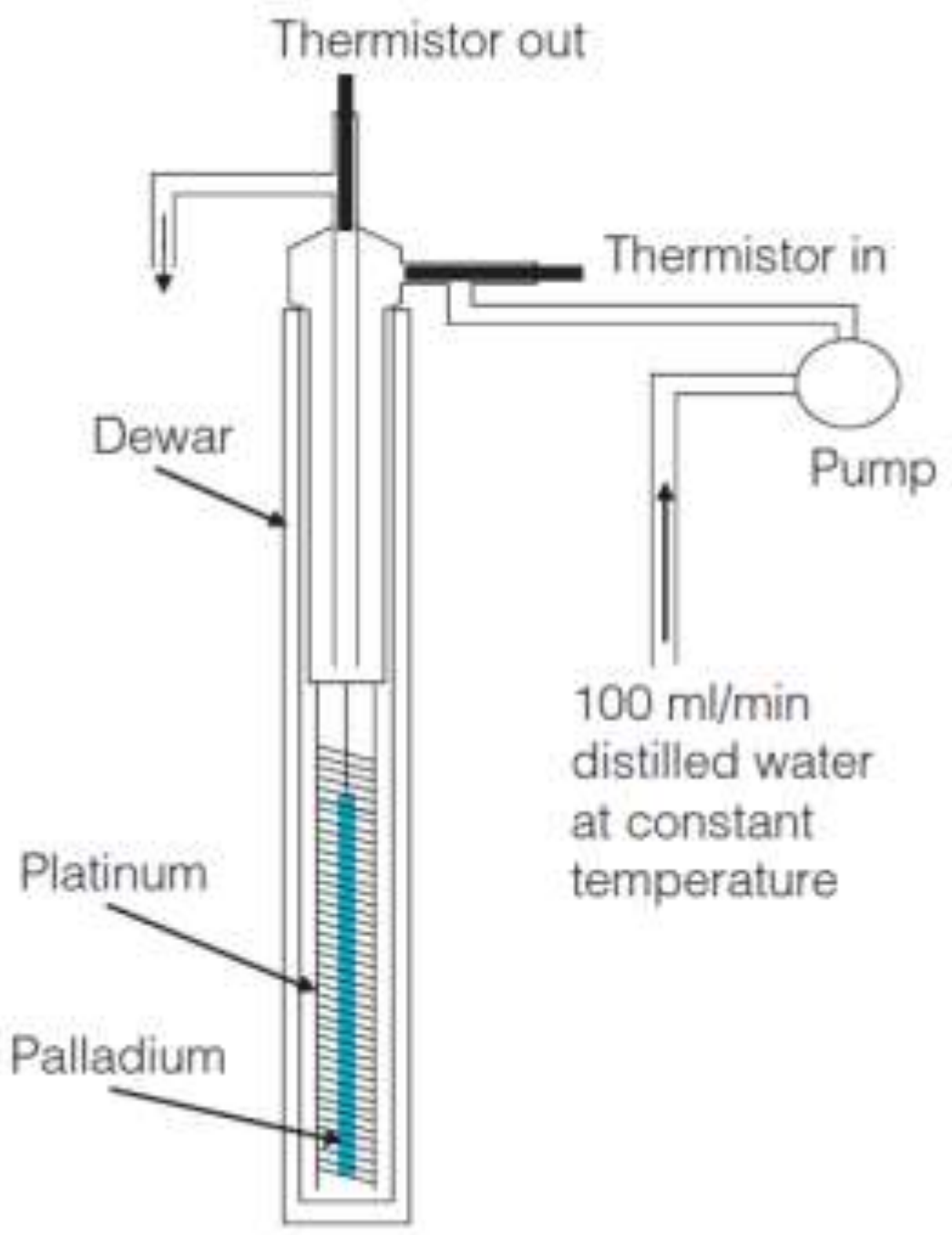

Figure 3. The test cell. [Biberian 2009]

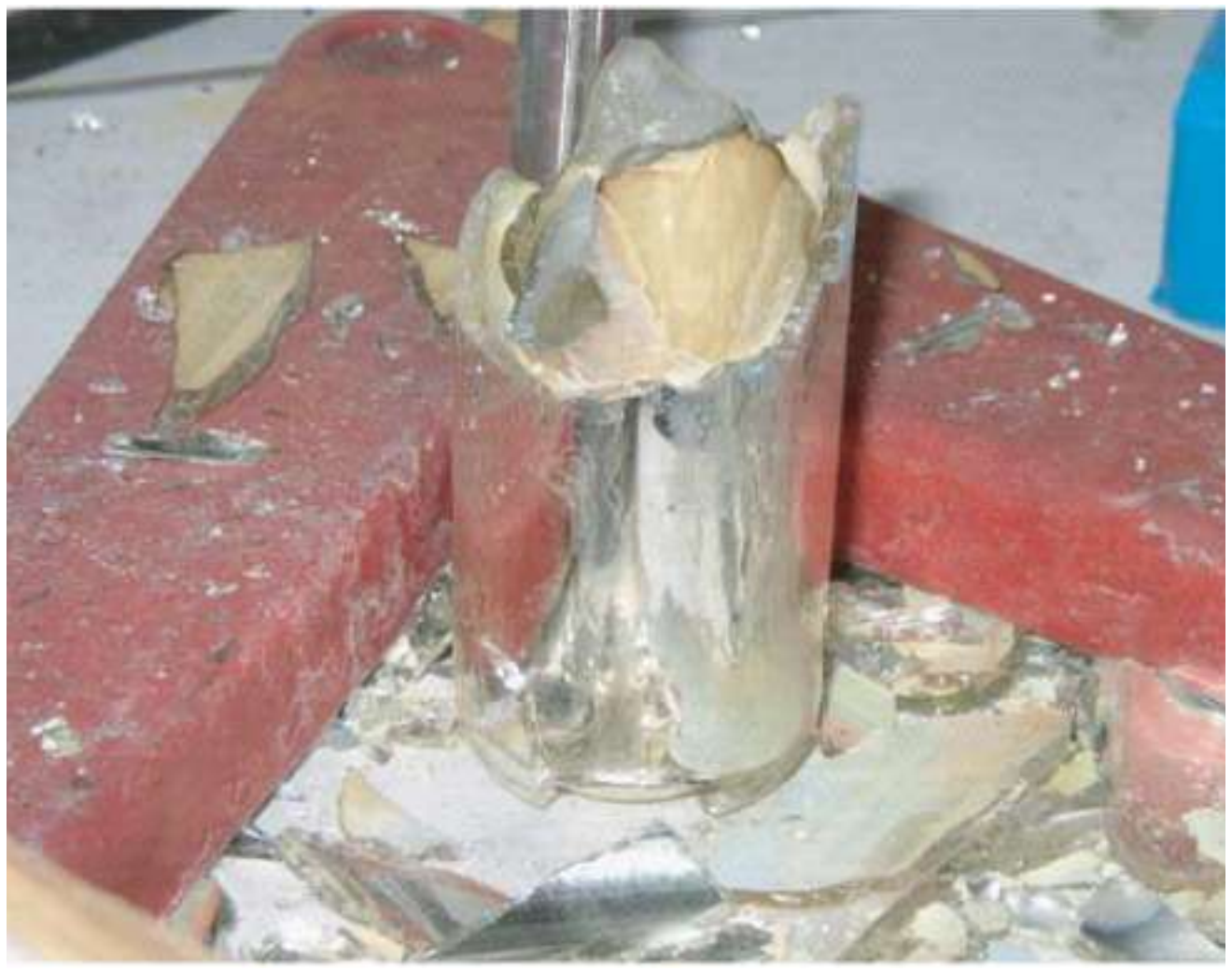

Figure 4. Remains of exploded dewar.

The author states, *"It is unlikely that the explosion was due to a deuterium oxygen recombination explosive reaction, since the cell was open, the amount of deuterium and oxygen gas was very limited in the cell and any pressure created by recombination should have escaped through the unsealed open end of the cell. It is very likely that under some not yet understood conditions, chain reactions occur in highly loaded palladium samples giving rise to an explosion.* [Biberian 2009]

To further test the chemical explosion hypothesis, a hydrogen-oxygen mixture was detonated in a similar cell, and the cell was not damaged.

- On January 24, 2005, Tadahiko **Mizuno** observed the explosion of a plasma electrolysis cell in Hokkaido, Japan. The water temperature rose from 25 C to 70 C in 10 s, and a bright glow gradually appeared at the bottom of the cathode, then expanded into the solution and exploded, Figure 5. Mizuno, about 1 m away, was cut by multiple glass shards and deafened for a week. [Mizuno 2005]

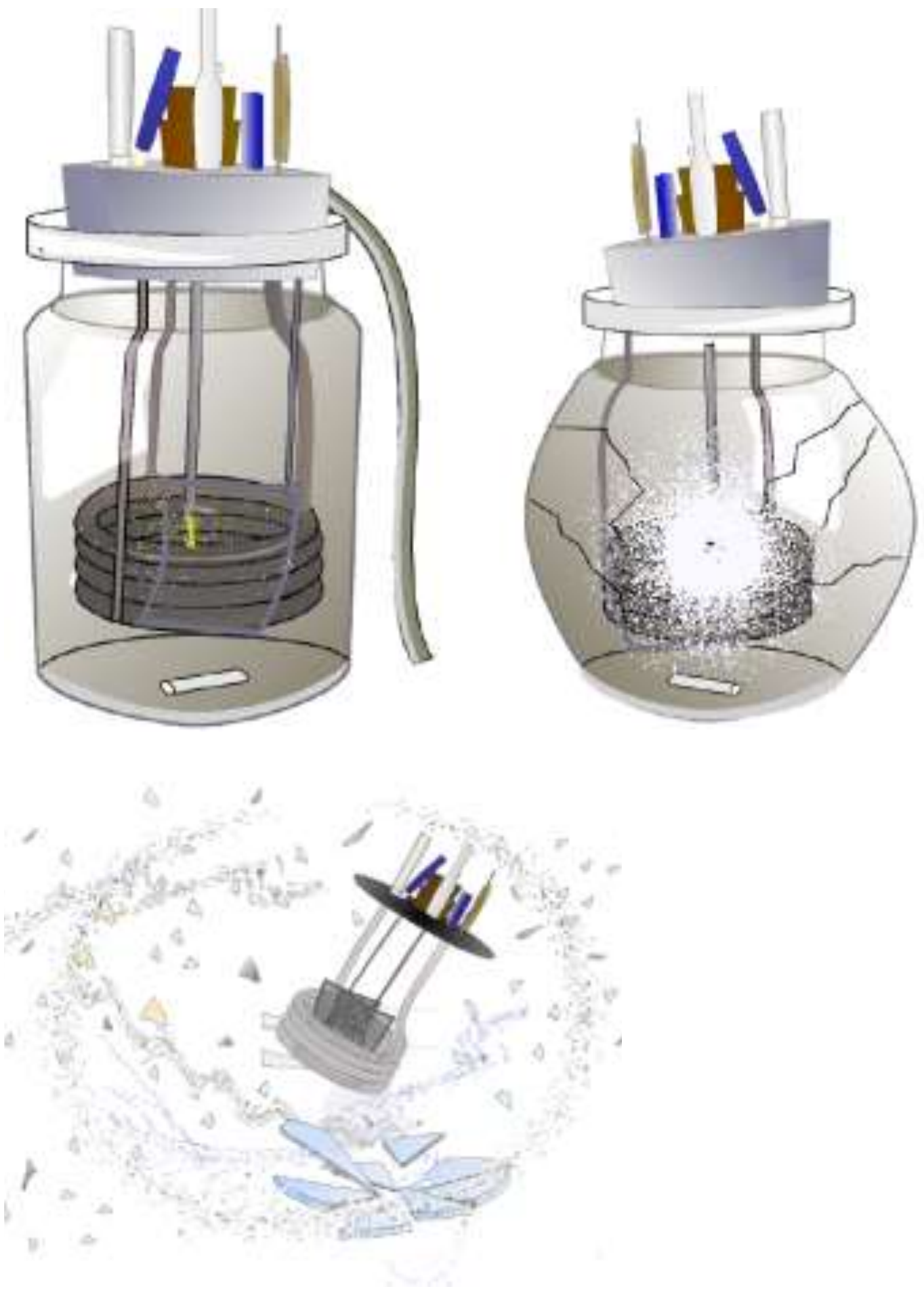

Figure 5. Illustration of Mizuno cell explosion. Glow begins below cathode tip (left), expands into solution in 10 s (middle), and vessel explodes.

The input power P ≈ 15 V x 1.5 A = 23 W for about 12 s ≈ 250 J, Figure 6. The slow growth of the output temperature probably rules out a hydrogen-oxygen detonation, and the output energy of the explosion was estimated to be about 800 times the input energy.

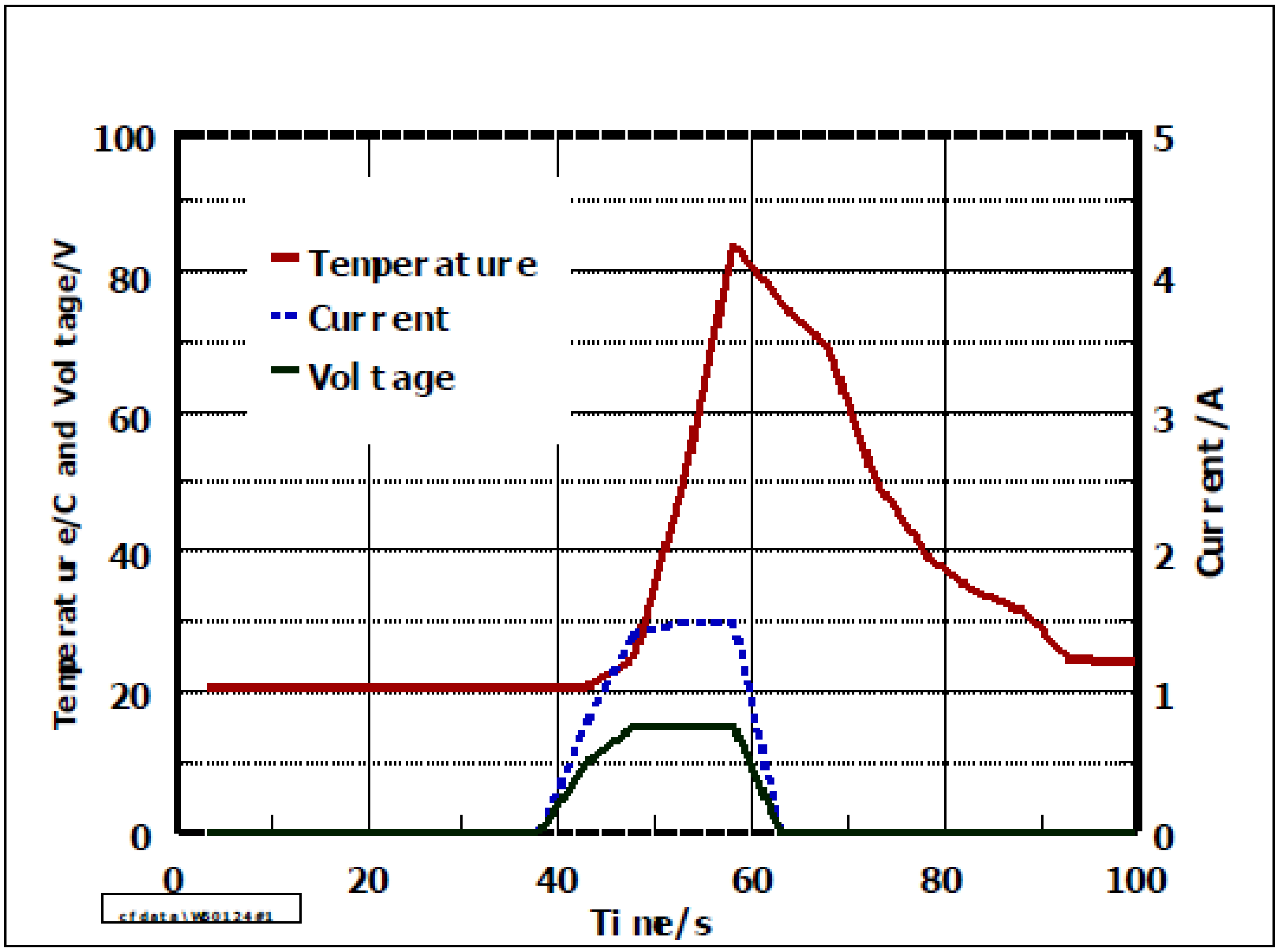

Figure 6. Input and output powers during the Mizuno explosion.

Many foreign elements (not originally present) were deposited on the tungsten electrode surface. The major detected elements were Si, S, Ca, Ti, Fe, Cr, and Cu.

## References to Appendix A